\documentclass[12pt]{article}
\usepackage{amsmath}
\usepackage{amssymb}
\usepackage{euscript}
\usepackage{cite}

\usepackage{alltt}

\begin{document}
\def\Was{W\c as}
\def\Order#1{${\cal O}(#1$)}
\def\lint{\int\limits}
\def\bbeta{\bar{\beta}}
\def\tbeta{\tilde{\beta}}
\def\talpha{\tilde{\alpha}}
\def\tomega{\tilde{\omega}}
\def\reff#1{(\ref{#1})}

\begin{titlepage}

\begin{flushright}
{\bf  CERN-TH/99-98  }
\end{flushright}

\vskip 2 cm

\begin{center}
{\bf\LARGE
How to Generate Four-Fermion Phase Space$\,^\dag$\\
}
\end{center}

\vskip 1truecm
%
\vskip 1truecm

\vspace{2mm}
\begin{center}
  {\bf   M. Skrzypek$^{a,b,\ddag}$}
  {\em and}
  {\bf   Z. W\c{a}s$^{a,b,\ddag}$ }
\\
\vspace{3mm}
{\em $^a$Institute of Nuclear Physics,
  ul. Kawiory 26a, 30-055 Cracow, Poland,}\\
{\em $^b$CERN, Theory Division, CH-1211 Geneva 23, Switzerland,}\\
\end{center}

\vspace{2cm}
\begin{abstract}

We present a scheme for integrating the matrix element of an arbitrary 
$e^+e^-\to f_1f_2\bar f_3\bar f_4$ process over the complete 
four-fermion phase space, or its any part, by means of the Monte Carlo technique.
The presented algorithm has been successfully implemented in the 
{\tt KORALW} Monte Carlo code.
\end{abstract}

\vspace{2.3cm}

\begin{center}
{\it To be submitted to Comput. Phys. Commun.}
\end{center}

\vspace{1.3cm}
\renewcommand{\baselinestretch}{0.1}
\footnoterule
\noindent
{\footnotesize
\begin{itemize}
\item[${\dag}$]
Work supported in part by 
Polish Government grants 
KBN 2P03B08414 and 
KBN 2P03B14715, and by 
the Maria Sk\l{}odowska-Curie Joint Fund II PAA/DOE-97-316.
\item[${\ddag}$]
Part of this study was performed while the authors were at the
{\em Inst. f\"ur Theoretische Physik, Karlsruhe Universit\"at, Karlsruhe,
Germany}, supported by 
a stipend within the EU grant no ERBCIPDCT940016.

\end{itemize}
}
\renewcommand{\baselinestretch}{1.0}

\vspace{0.1cm}
\begin{flushleft}
{\bf CERN-TH/99-98 \\ April 1999}
\end{flushleft}

\end{titlepage}


\newpage

\section{Introduction}


The measurement of the different observables related to the so-called
four-fermion processes constitutes an important part of LEP2 physics.
The most important group of observables is related to $W$-pair production 
and decay. It is not only the total cross section we are interested in, 
but also the wealth of different spin and angular asymmetries. This leads 
to rather sophisticated observables depending on multidimensional distributions 
that are also often strongly affected by experimental cut-offs.
There are {\it nine} distinct $W$ decay channels and, owing to small statistics
of $W$-pair production, one is often bound to combine data of rather different 
experimental signatures to obtain statistically meaningful results.
The situation of the forthcoming $ZZ$ physics or so-called single-$W$ physics
is rather similar. In all cases, theoretical predictions, to be comparable 
with the data, have to be provided in the form of
four-fermion Monte Carlo simulations. 

Monte Carlo integration over four-body
phase space%
\footnote{
 In reality the case is even more complicated, because of 
 additional bremsstrahlung photons. In this paper, however,
 we restrict ourselves
 to pure Born-level phase space and {\it do not} discuss any
 issues related to photonic corrections. Let us note, only, that in the general
case the Born-level phase space can form well-defined building-blocks of the
whole algorithm.
}
 is a non-trivial problem. Why is it so?
%
First of all, there is over a hundred different four-fermion 
final states allowed within the Standard Model. They differ by the flavours
of final-state quarks and/or leptons. Each of these states, even at the Born
level, is described by many, often over a hundred, Feynman graphs.
For distinct four-fermion final states one finds thus diametrically different
structures of singularities
depending on the set of Feynman graphs describing the particular 
process. The singularities, or more precisely sharp peaks, have to be 
carefully tackled by the generators to assure their efficiency. They can
affect the distribution in one or more variables of the particular 
parametrization of the phase space.
As an example we can think of the $t$-channel singularities $1/t$, 
$s$-channel ones such as resonances, etc.
These singularities occur in various parts of the phase space. Also, if they
are present in more than one variable
they can affect one another, either destructively or by
increasing the degree of singularity of the second one. Finally one particular 
four-fermion process can have many distinct singular sub-processes, which need
to be treated individually.

Of course, there are also some similarities and symmetries amongst 
processes as well as individual Feynman graphs. If properly used, they 
can somewhat reduce the size of the problem, but unfortunately its complexity 
remains intact. At the same time, if one departs from the Standard
Model and introduces any new particles or interaction types, the
size and complexity of the problem grows. 

At the top of the above list of difficulties, there is also a strong cut-off 
dependence of the cross-section. This, in the case 
of electrons close to the beam directions,
lead either to instability of the Monte Carlo integration 
or makes the generation ineffective since
many events are generated outside the detector acceptance.

In the present paper we show a working solution to the above phase-space 
integration problem. It is based on the
{\it multichannel} approach that is  
used in the {\tt KORALW} Monte Carlo code \cite{koralw:1995a,koralw:1998}.
Our aim was to enable the generation of all four-fermion  
$e^+e^-\to f_1f_2\bar f_3\bar f_4$ final states
of the Standard Model over the full phase space or its  sub-space,
within reasonable CPU time.

At the time of the LEP2 Workshop \cite{yr:lep2},
several dedicated four-fermion Monte Carlo codes were presented. 
As far as the integration method is
concerned, some general strategies can be identified amongst these
codes:  
the adaptive
algorithms based on {\tt VEGAS}-type routines
\cite{alpha,wphact,wwf,wwgenpv},  the {\tt BASES-SPRING}
algorithm \cite{grc4f,comphep,wwf}, the importance-sampling technique  
\cite{wwgenpv,lpww02,wopper} 
and the {\it multichannel} importance-sampling algorithms.
In the latter group, one finds two approaches: of the {\tt EXCALIBUR} code
\cite{berends:1994,excalibur:1995}, followed by {\tt ERATO}
\cite{erato:1997}, and of the {\tt KORALW} \cite{koralw:1995a,koralw:1998},
the subject of this paper.

The major advantage of the {\it multichannel} approach is, in our opinion, 
that it gives almost complete control over the presampling process. 
The branches correspond to physical configurations in a transparent
way. At the same time, since one can implement many branches, 
the structure of the generator can be very rich. 
No complicated dividing of the multidimensional phase space is
necessary and no {\it potentially uncontrolled} automated optimization 
in the core of 
the algorithm needs to be performed (although some {\em auxiliary}
optimization at the top levels can be useful).

The {\it multichannel} approach to few-body phase-space integration 
has been used already in the past. For instance,
the {\tt TAUOLA} \cite{tauola:1990,tauola:1992,tauola:1993} 
package for $\tau$ decays works in such a manner up to five-body decays. The 
{\tt FERMISV} \cite{kleiss:1993} 
code generates the NC-type four-fermion processes in that way also.
As it happens with the Monte Carlo algorithms, it may be difficult
to point to the exact source of the {\it multichannel} concept. Traces 
of it can be found in a number of 
other papers as well. Let us mention
just a few of these: \cite{fowl,mustraal,jadach:1985,topki,zodec}.

It is instructive to note that
there are two ``layers'' in the structure of {\it multichannel} 
algorithms. The
core of the algorithm is determined by the internal structure of the 
{\it channels} that contain the process-specific information on the 
actual generation
algorithm. However, on the top of separate channels there is another,
external layer, 
responsible for the ``branching'' process itself. 
Recently \cite{jadach:MC-www} this second step has
been explained in detail. 

The main advantage of the algorithm presented here is, in our opinion, the
following. Out of various, already known
as well as new, ideas, we show how to create in a systematic way,
with simple building blocks, a
universal and extendable presampler for four-body final states.
Despite its simplicity, this presampler is shown to work effectively 
in an actual Monte Carlo code {\tt KORALW}. The important advantage of 
the algorithm proposed here is its
modularity and simplicity: all the branches are built out of two
rather simple elementary pieces.

In this paper we present a complete description of the multichannel
four-fermion Monte Carlo algorithm of the {\tt KORALW}
code. 
We start with the detailed description  of the ``external layer'' -- 
that is the branching process -- which is quite universal. 
Using notation of ref.\ \cite{jadach:MC-www}, we show
how it works in the existing algorithm of {\tt KORALW}. Starting from 
section\ 3, we discuss the ``inner layer'' -- the actual
structure of individual channels dedicated to generating the four-fermion
phase space. 
We end the paper with a short summary.


\section{``Multichannel'' Master Formula}
\label{MASTER}

As a basis of the generation of the four-fermion phase space we take
a multibranch type of Monte Carlo algorithm. 
In this section, let us  summarize the details of the ``external layer''
of four-fermion phase space.
Our goal is to calculate the following integral
\begin{eqnarray}
\label{SIGMA}
\sigma &=& \int d\Phi(P,\underline q) \vert M(P,\underline q) \vert^2 
\\
d\Phi(P,\underline q) 
    &=&\frac{1}{2P^2} (2\pi)^4\delta^4\left(P-\sum_{k=1}^r q_k\right)
           \frac{1}{2^{r-1}} 
           \prod_{k=1}^r \frac{1}{(2\pi)^3}\frac{d^3\vec{q_k}}{2E_k},
           \;\; r=4,
\label{PHASE_SP}
\end{eqnarray}
where $P$ denotes the sum of beam momenta, $q_k$ denote the four-momenta of
final particles, and $\underline q$ denotes collectively all $q_k$
four-momenta.  
Note that eq.\ \reff{PHASE_SP} 
includes all the normalization
factors, including the flux factor $1/2P^2$. 


Our strategy in solving eq.\ \reff{SIGMA} is to simplify $\vert M\vert^2$
until it reaches the analytically calculable level. Then, we generate
this simple distribution and reweight it to an exact $\vert M\vert^2$.

Let us then replace $\vert M\vert^2$ with the crude distribution 
$\tilde\psi_{CR}$, in which we introduce the
structure of branches
\begin{equation}
d\Phi(P,\underline q) \tilde \psi_{CR}
 =  d\Phi(P,\underline q) \sum_i^{N_{BR}} \tilde p_i 
        \frac{\tilde \psi_{CR}^i(P,\underline q)}
             {{\cal J}^i(P,\underline q)}.
\label{TREE}
\end{equation}
The summation goes over branches, with $N_{BR}$ denoting the total number of
branches. 
In this equation we introduced two set of functions: $\tilde \psi_{CR}^i$
and ${\cal J}^i$. The first
represents the actual (crude) distributions to be generated over the phase space.
It is in these functions that all the physical information is
located. Their specific form is the main subject of this paper and will
be analysed in detail in the next sections. 
Here we confine ourselves to the formal structure of
the arrangement of branches.
${\cal J}^i$ is composed of
the Jacobians of the change 
of variables from four-momenta to angular and invariant-mass variables
(as defined in formulae below), which we anticipate in
each channel $i$ and of which we want to get rid (and reintroduce later by
appropriate reweighting). They are isolated in eq.\ \reff{TREE} for
simplicity reasons.
Finally, $\tilde p_i$ are some positive coefficients to be specified later.
We change variables $(P,\underline q)$ into
angles and masses 
$(\cos\underline \theta^i,\underline \phi^i,\underline s^i)$ 
defined for each branch in different Lorentz frames. By 
$\cos\underline \theta^i,\underline \phi^i,\underline s^i$ 
we denote collectively all the angles and masses of a given branch $i$.
The crude distribution becomes
\begin{equation}
\sum_i^{N_{BR}} 
d\Phi^i(\cos\underline \theta^i,\underline \phi^i,\underline s^i) 
\tilde p_i 
\tilde \psi_{CR}^i(\cos\underline \theta^i,\underline \phi^i,
                                                           \underline s^i), 
\label{TREE-BRANCH}
\end{equation}
where we introduced the explicit form of the Jacobian
${\cal J}^i(\cos\underline \theta^i,\underline \phi^i,\underline s^i)$ 
in new variables
\begin{eqnarray}
{{\cal J}^i(\cos\underline \theta^i,\underline \phi^i,\underline s^i)} 
  &=& \frac{d\Phi(P,\underline q)}
         {d\Phi^i(\cos\underline \theta^i,\underline \phi^i,\underline s^i)}
  = \prod_{j=1}^3 \lambda_j^i,
\label{TREE-JACOBIAN}
\\
\lambda_j^i=\lambda(1,s_{2_j}^i/s_{1_j}^i,s_{3_j}^i/s_{1_j}^i) &=&
\frac{1}{s_{1_j}^i} 
     \sqrt{ \left( s_{1_j}^i -s_{2_j}^i -s_{3_j}^i \right)^2
       -4 s_{2_j}^i s_{3_j}^i},
\label{TREE-LAMBDA}
\\
d\Phi^i (\cos\underline \theta^i,\underline \phi^i,\underline s^i)
      &=& \frac{1}{4P^2} \frac{1}{(4\pi)^{8}}
       ds_1^ids_2^i \prod_{j=1}^{3} d\cos\theta_j^i d\phi_j^i.
\end{eqnarray}

The mass-like variables $s_1^i$ and $s_2^i$ present in the integration element
denote squares of the invariant masses of systems consisting of two or three
outgoing fermions. They are present in the definition of the $\lambda$-factors 
of the $\cal J$ as well, but in this case $s^i_{j_k}$
can denote squares of  masses of the outgoing fermions as well. 

We have found that
it is convenient to enlarge integration domains of the branches. 
This can be done for each branch in a different way. We do it by 
introducing $\Theta_i$ functions
\begin{equation}
\label{THETA}
\tilde \psi_{CR}^i(\cos\underline \theta^i,\underline \phi^i,
\underline s^i) =
\psi_{CR}^i(\cos\underline \theta^i,\underline \phi^i,\underline s^i) 
\Theta_i(\Omega^i)
\end{equation}
and extending the integration areas
$\Omega^i \to \Omega^i_{CR}$.
The $\Theta_i$ enforce the exact phase-space limits. Now, by setting
$\Theta_i\to 1$, we arrive at the final form of the crude distribution
\begin{equation}
\sum_i^{N_{BR}} 
d\Phi^i(\cos\underline \theta^i,\underline \phi^i,\underline s^i)
 \vert_{\Omega^i_{CR}}
\tilde p_i \psi_{CR}^i(\cos\underline \theta^i,\underline \phi^i,
                                                           \underline s^i). 
\label{TREE-CRUDE}
\end{equation}
 We assume that eq.\ \reff{TREE-CRUDE}
is integrable analytically, with the result 
\begin{equation}
\int\limits_{\Omega^i_{CR}} \tilde p_i 
  d\Phi^i(\cos\underline \theta^i,\underline \phi^i,\underline s^i)
\psi_{CR}^i(\cos\underline \theta^i,\underline \phi^i,
                                                           \underline s^i) 
 = \tilde p_i{\cal N}_{CR}^i.
\label{TREE-NORM}
\end{equation}
The random choice of the branch to be used in the generation of a particular
point in the phase space is performed with the help of probabilities $P_i$:
\begin{equation} 
\label{PI}
P_i=\frac{\tilde p_i{\cal N}_{CR}^i}
         {\sum\limits_k^{N_{BR}} \tilde p_k {\cal N}_{CR}^k}.
\end{equation}
Later, following ref.\ \cite{jadach:MC-www}, we generate points
according to distributions $\psi_{CR}^i$ of eq.\ \reff{TREE-CRUDE}. 
The last step to be done is to reintroduce all the simplifications in
the form of the appropriate weights to complete the prescription on how to
calculate  the integral \reff{SIGMA}.
For each channel $i$, the transition from $\psi_{CR}^i$ of eq.\
\reff{TREE-CRUDE} to $\tilde \psi_{CR}^i$ of eq.\ \reff{TREE-BRANCH}
can be achieved with the help of a simple step-like weight: 
\begin{equation} 
w^a(i)=\frac{\tilde\psi_{CR}^i}{\psi_{CR}^i}=\Theta_i. 
\label{WA}
\end{equation} 
The whole distribution of eq.\ \reff{TREE-BRANCH} is then generated from 
$\psi_{CR}^i$ with the help of the same weight $w^a(i)$, the argument $i$
being the number of the chosen branch for a particular event. The
distribution generated this way 
\begin{equation}
\sum\limits_i^{N_{BR}} P_i \frac{\psi_{CR}^i}
                                  {\tilde p_i {\cal N}_{CR}^i} w^a(i)
= \frac{\sum\limits_i^{N_{BR}}\tilde p_i\tilde\psi_{CR}^i}
       {\sum\limits_i^{N_{BR}}\tilde p_i {\cal N}_{CR}^i}
\end{equation}
is, up to a normalization factor, equal to that of eq.\
\reff{TREE-BRANCH}. 
The normalization (integral) of eq.\ \reff{TREE-BRANCH} can
be calculated as
\begin{eqnarray}
\tilde{\cal N}_{CR}
&=& \sum\limits_i^{N_{BR}}\tilde p_i \tilde {\cal N}^i_{CR}
= \sum\limits_i^{N_{BR}}\tilde p_i {\cal N}^i_{CR} 
\left<w^a(i)\right>_i
\\
&=& \left(\sum\limits_k^{N_{BR}}\tilde p_k {\cal N}^k_{CR} \right)
  \left(\sum\limits_i^{N_{BR}} P_i \left<w^a(i)\right>_i \right)
= \; \Bigl<_{P_i}\left<w^a(i)\right>_i \Bigr> 
\sum\limits_k^{N_{BR}}\tilde p_k {\cal N}^k_{CR},
\end{eqnarray}
where we have made use of the fact that $P_i$ defines a ratio of 
the number of events generated
in channel $i$ to the total number of events. With $\left<x \right>_i$ we
denote the 
average of $x$ within the  $i$-th channel and with 
$\left<_{P_i} \; ... \; \right>$ we denote consecutive 
average with respect to the branches:
\begin{equation}
\Bigl<_{P_i} \left<w(i,\dots)\right>_i \Bigr>=
\sum\limits_i^{N_{BR}} \frac{N_i}{N} \left<w(i,\dots)\right>_i
\; ; \; 
\lim_{N \to \infty }
\Bigl<_{P_i} \left<w(i,\dots)\right>_i \Bigr>=
\sum\limits_i^{N_{BR}} P_i \left<w(i,\dots)\right>_i. 
\label{szesnascie}
\end{equation}
In practice, the external summation over channels is realized
with the help of the Monte Carlo generation 
of the channel number with the probabilities $P_i$ 
and, for every event, only that individual weight of the particular 
channel is calculated and used.
Furthermore, we do not need to
calculate $\Bigl<_{P_i} \left<w(i,\dots)\right>_i \Bigr>$
as an explicit sum of averages $\left<w(i,\dots)\right>_i$, calculated
and stored
for each channel separately, as in eq.\ \reff{szesnascie}.
Instead we  calculate only {\it one} average over the whole sample
of generated events of  all channels:
\begin{equation}
\Bigl<_{P_i} \left<w(i,\dots)\right>_i \Bigr>
= \sum\limits_i^{N_{BR}} \frac{N_i}{N} \frac{1}{N_i}
\sum\limits_{events~in~i}^{N_{i}} w(i,\dots)
=  \frac{1}{N} \sum\limits_{all~events}^{N} w(i,\dots)
\equiv \left< w(i,\dots)\right>.
\end{equation}
Here the argument $i$ denotes that for the  event generated
with channel $i$ we calculate a weight $w(i,\dots)$.
%
\footnote{
Let us also note that in general, in our paper,
we will omit the symbol $\lim_{N\to \infty}$.}

Including the fundamental matrix-element weight 
\begin{equation}
w^b(P,\underline q)
  =\frac{\vert M(P,\underline q)\vert^2}{\tilde\psi_{CR}(P,\underline q)}
\end{equation}
and repeating the above steps
we end up with the following formula for the integral \reff{SIGMA}
\begin{eqnarray}
\label{SIGMA2}
\int d\Phi(P,\underline q) \vert M(P,\underline q) \vert^2  
&=& 
\left< \tilde w(k,P,\underline q) \right>
\sum\limits_i^{N_{BR}}\tilde p_i {\cal N}^i_{CR},
\\
\tilde w(k,P,\underline q) &=& w^a(k) w^b(P,\underline q)
= 
\frac{\vert M(P,\underline q)\vert^2}{\tilde\psi_{CR}(P,\underline q)}
  \Theta_k(\Omega_k).
\end{eqnarray}
Let us concentrate for a moment on the coefficients $\tilde p_i$. 
If we rewrite them in
a more specific form as $\tilde p_i =p_i/{\cal N}^i_{CR}$, with the
condition $\sum p_i =1$, we find from eq.\ \reff{PI} that $P_i = p_i$, so
that the 
$p_i$ have a nice interpretation as the actual branching probabilities.
The master equation \reff{SIGMA2} then evolves into
\begin{eqnarray}
\label{SIGMA3}
\int d\Phi(P,\underline q) \vert M(P,\underline q) \vert^2
&=&   
\left<w(k,P,\underline q)\right>,
\\
w(k,P,\underline q) &=& 
\vert M(P,\underline q)\vert^2
    \left[ \sum\limits_i^{N_{BR}} 
        \frac{p_i \tilde \psi_{CR}^i(P,\underline q)}
             {{\cal N}^i_{CR}{\cal J}^i(P,\underline q)}
    \right]^{-1}
\Theta_k(\Omega_k).
\end{eqnarray}
Much as in  ref.\ \cite{jadach:MC-www}, we will now discuss 
a possible variation of the above algorithm. Turning back to
 eq.\ \reff{TREE},
we modify it by removing the Jacobians ${\cal J}_i$:
\begin{equation}
d\Phi(P,\underline q) \tilde \psi_{CR}
 =  d\Phi(P,\underline q) \sum_i^{N_{BR}} \tilde p_i 
        \tilde \psi_{CR}^i(P,\underline q)
\label{TREE-BIS}
\end{equation}
and instead introduce them in eq.\ \reff{TREE-BRANCH}
\begin{equation}
\sum_i^{N_{BR}} 
d\Phi^i(\cos\underline \theta^i,\underline \phi^i,\underline s^i) 
\tilde p_i 
{\cal J}^i(\cos\underline \theta^i,\underline \phi^i,\underline s^i)
\tilde \psi_{CR}^i(\cos\underline \theta^i,\underline \phi^i,
                                                           \underline s^i). 
\label{TREE-BRANCH-BIS}
\end{equation}
On the course to eq.\ \reff{TREE-CRUDE} we simplify 
${\cal J}^i_{CR} \to 1$ and, as a result, weights $w^a(i)$ get modified:
\begin{equation} 
w^a(i)=\frac{{\cal J}^i\tilde\psi_{CR}^i}{\psi_{CR}^i}
     ={\cal J}^i\Theta_i. 
\label{WA-BIS}
\end{equation} 
Accordingly, eq.\ \reff{SIGMA3} becomes
\begin{eqnarray}
\label{SIGMA3-BIS}
\int d\Phi(P,\underline q) \vert M(P,\underline q) \vert^2  &=&
\left<w(k,P,\underline q)\right>,
\\
w(k,P,\underline q) &=& 
\vert M(P,\underline q)\vert^2
    \left[ \sum\limits_i^{N_{BR}} 
        \frac{p_i \tilde \psi_{CR}^i(P,\underline q)}
             {{\cal N}^i_{CR}}
    \right]^{-1}
{\cal J}^k(P,\underline q)\Theta_k(\Omega_k).
\end{eqnarray}

Note that here the Jacobians $\cal J$ are {\em not} summed over branches but
calculated for the branch of a given event only. As the actual form of
the Jacobians is rather simple, we should not expect a significant
difference in the performance of the two algorithms.

Finally, let us remark that {\em both} of our algorithms are in fact of
the ``second'' type in the classification of
ref. \cite{jadach:MC-www}. 
In the first one (eq.\ \reff{SIGMA3}), we attribute zero weight to some 
events already inside branches in the form of trivial step
functions $\Theta_i$. In the second algorithm, 
weights $\Theta_i {\cal J}^i$ are continuous functions between zero and one. 

In the language of measure theory the difference of the two algorithms 
(defined respectively 
by formulae \reff{SIGMA3}
and \reff{SIGMA3-BIS} ) lies in the choice of the basic measure on the 
phase-space
manifold. In the second, more natural case, the Lorentz-invariant phase-space
element takes this role.

\section{The Branches}
\label{BRANCHES}

In the previous section we 
gave the complete general description of the
``external layer'' of the multichannel algorithm responsible for the
alignment of separate branches. In this section we will present the
construction of the actual branches in the case of four-fermion final
states that we have developed for the {\tt KORALW} code \cite{koralw:1998}.

Let us begin with remark. 
Each branch $\psi^i_{CR}$ of eqs.\
\reff{TREE-BRANCH}, \reff{TREE-CRUDE}, \reff{TREE-BRANCH-BIS} 
is intended to describe a certain type of
singularities that one encounters in the Feynman graphs,
but there is no unique definition of 
the $\psi^i_{CR}$ functions. They are dummy functions that cancel out in
the final result. On the one hand, their role is to mimic as close as
possible the
complicated structure of singularities of the true matrix element, but
on the other hand they should not be too complicated. First of
all they {\it must}
be analytically integrable over the generation area in order to be able
to normalize the generator. 
Secondly, because of the complexity of the problem, it is, in our
opinion, very useful for them to
have a modular structure. This allows us
to write the program and its documentation in a compact and
transparent way, leaving room for easier modifications in the
future.  
Finally, these simplifications obviously cannot go too far
as the $\psi^i_{CR}$ functions have to be quite universal and flexible
to accommodate a variety of matrix-element singularities.

How do we fulfil in practice all these, partly contradictory, requirements?
How do we construct the actual $\psi^i_{CR}$ functions? 


The crucial assumption that we make 
is the partial 
factorization of $\psi^i_{CR}$ with respect to the angular and mass 
variables (we skip the branch index $i$ for the next two sub-sections):
\begin{equation}
\psi_{CR}(\cos\underline \theta,\underline \phi,\underline s)
  =f(\underline s)g(\cos\underline \theta,\underline \phi,\underline s).
\label{FACTORISATION}
\end{equation}
We request that 
$g(\cos\underline \theta,\underline \phi,\underline s)$, upon integrating 
over angular variables, becomes independent of masses $\underline s$
\begin{equation}
\int d\cos\underline \theta d\underline \phi 
   g(\cos\underline \theta,\underline \phi,\underline s)
   = \hbox{const}.    
\end{equation}
Finally, we assume that masses $\underline s$ will always be
generated first, before the angles $\underline \theta,\underline
\phi$. 

With all these simplifications, one may worry that we are too
restrictive. For example, it may happen that, for certain configurations
of singularities, it would be natural to reverse the order of generation --
angles first and then masses, or even, further, that one should use some
combination of these variables as more ``physical''. 
This approach would, however, betray to some extent the modularity of the
program: it is always up to the author of an algorithm to decide where
to draw the line.
Our priority was set on simplicity and modularity of the algorithm.
We believe we achieved those to a large extent. 
The solution presented here may look 
trivial and simplistic, but the point is
that the efficiency of the generation
is sufficient for our purposes, and we do not need to pay
the price of formulas several pages long, which may be necessary 
e.g. in the 
algorithm of refs.\ 
\cite{kleiss:1993,excalibur:1995}.

Now we can proceed to the details
of the mass and angular distributions $f$ and
$g$.


\subsection{Mass Distributions}


As explained above we intend to
generate $s_1, s_2$ variables first, and then the angles. 
Generically there are two possible ways of aligning the two mass
variables: a ``chain''-like and a ``fork''-like, see Fig.\ 1.

\noindent
\unitlength=1mm
\begin{picture}(140,70)(0,0)
\put(10,40){\line(1,0){60}}
\put(25,40){\line(1,-1){10}}
\put(40,40){\line(1,-1){10}}
\put(55,40){\line(1,-1){10}}

\put(10,42){$s_{max}$}
\put(30,42){$s_{1}$}
\put(45,42){$s_{2}$}
\put(65,42){$m_{4}$}
\put(65,26){$m_{3}$}
\put(50,26){$m_{2}$}
\put(35,26){$m_{1}$}

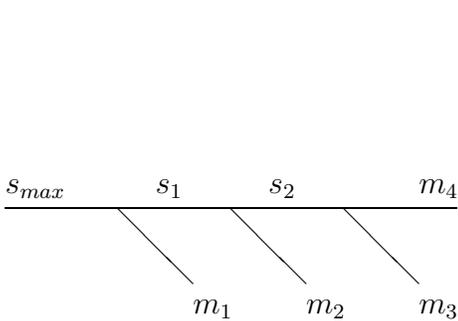
\put(10,5){Fig.\ 1a: ``Chain''}


\put(85,40){\line(1,0){15}}
\put(100,40){\line(1,1){25}}
\put(100,40){\line(1,-1){25}}
\put(115,55){\line(1,-1){10}}
\put(115,25){\line(1,1){10}}

\put(85,42){$s_{max}$}
\put(100,47){$s_{1}$}
\put(100,32){$s_{2}$}
\put(127,13){$m_{4}$}
\put(127,35){$m_{3}$}
\put(127,43){$m_{2}$}
\put(127,65){$m_{1}$}

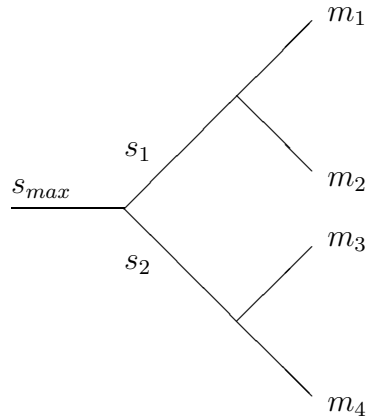
\put(100,5){Fig.\ 1b: ``Fork''}

\end{picture}

For each of these configurations we need to give the exact phase-space
limits $\Theta(\Omega)$ corresponding to eq.\ \reff{THETA}. 
Then we define an extension 
$\Omega \to \Omega_{CR}$
of these limits that enable the analytic integration of the
distributions to simple results, allowing
numerically fast normalization to unity.

For the two cases defined above we continue independently:

\noindent
\unitlength=1mm
\begin{picture}(150,80)(0,0)
\label{STRIPES}

\put(0,0){
\begin{picture}(75,80)(-10,-10)

\put(10,10){\framebox(40,40)}

\put(10,10){\line(1,1){40}}

\put(35,0) {\large $s_{1}$}
\put(-5,35){\large $s_{2}$}

\put(47,6){$s_{max}$}
\put(1,50){$s_{max}$}
\put(6,6){$0$}

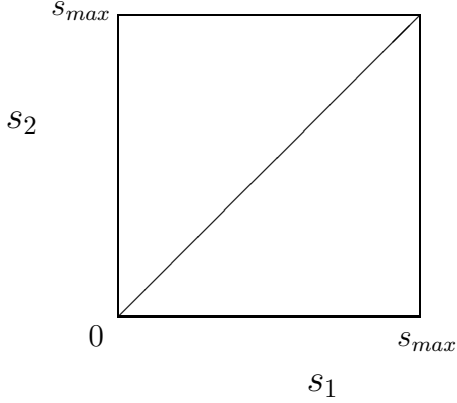
\put(-5,-10){Fig.\ 2a: ``Chain''-type phase space}

\end{picture}
}

\put(80,0){
\begin{picture}(75,80)(-10,-10)

\put(10,10){\framebox(40,40)}
\qbezier(50,10)(10,10)(10,50)

\put(10,20){\line(1,0){1}}
\put(20,10){\line(0,1){1}}

\put(20,20){\line(1,0){30}}
\put(20,20){\line(0,1){30}}

\put(35,0) {\large $s_{1}$}
\put(-5,35){\large $s_{2}$}

\put(47,6){$s_{max}$}
\put(1,50){$s_{max}$}
\put(17,6){$\frac{s_{max}}{4}$}
\put(1,20){$\frac{s_{max}}{4}$}
\put(6,6){$0$}

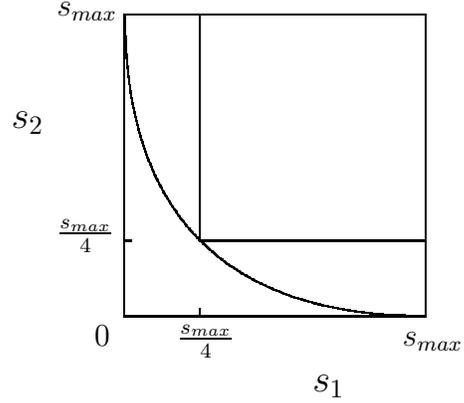
\put(-5,-10){Fig.\ 2b: ``Fork''-type phase space}

\end{picture}
}

\end{picture}

\begin{itemize}
\item
The ``chain''.\\
The exact phase-space limits $\Theta(\Omega^C)$ of eq.\ \reff{THETA} 
in this case are given by:
\begin{equation}
\sqrt{s_{max}} \geq \sqrt{s_1}+m_1, \;\;
\sqrt{s_1} \geq \sqrt{s_2}+m_2,\;\;
\sqrt{s_2} \geq m_3+m_4.
\end{equation}
By dropping masses, the above domain $\Omega^C$ extends 
to the simpler one,  $\Omega^C_{CR}$ of\ Fig
2a:
\begin{equation}
s_{max} \geq {s_1} \geq {s_2} \geq 0.
\end{equation}
Next, we assume that $s_1$ and $s_2$ distributions are identical and that
the function $f(s_1,s_2)$ factorizes 
\begin{equation}
f(s_1,s_2)= \frac{1}{{\cal F}(s_{max})}f_1(s_1) f_1(s_2).
\end{equation}
This is a somewhat restrictive assumption, but as there will be practically
no restrictions on the form of a single distribution $f_1(s_1)$, 
we can always use a sufficiently general form of it. 
It costs some efficiency, but this is
compensated by the (almost) perfect representation of the phase-space limits
already at a crude level and by the speed of the code. 

With our assumptions, the integral 
${\cal F}$ can be easily calculated:
\begin{eqnarray}
{\cal F}(s_{max}) &=& \int_{\Omega^C_{CR}} f(s_1,s_2)ds_1 ds_2 = 
\int_0^{s_{max}} ds_1  f_1(s_1)
\int_0^{s_1} ds_2  f_1(s_2)
 \\
&=& \frac{1}{2}\left[ F_1(s_{max})- F_1(0)\right]^2,
\\
 F_1(x) &=& \int^x  f_1(s)ds
\end{eqnarray}
and the distribution can be normalized, assuming one knows the $F_1$
function.
%
\item
The ``fork''.\\
Also in this configuration we assume factorization of the function 
$f(s_1,s_2)$ into the one-dimensional functions $f_1(s_1)$ and  $f_2(s_2)$:
\begin{equation}
f(s_1,s_2)= \frac{1}{{\cal F}(s_{max})}f_1(s_1) f_2(s_2).
\end{equation}
The phase-space limits 
($\Theta(\Omega^F)$ of eq.\ \reff{THETA}) are the following:
\begin{equation}
\sqrt{s_1} +\sqrt{s_2} \leq \sqrt{s_{max}},\;\;
\sqrt{s_1} \geq m_1 +m_2,\;\;
\sqrt{s_2} \geq m_3 +m_4.
\end{equation}
This area
can be extended to the polygonal $\Omega^F_{CR}$ domain, or explicitly 
to the surface defined by the difference between the two squares, as shown in 
Fig.\ 2b.  
The integration is easy: 
\begin{eqnarray}
{\cal F}(s_{max}) &=&
\int_{\Omega^F_{CR}} f_1(s_1)f_2(s_2)ds_1 ds_2 
 \\
&=&
\left[\int_0^{s_{max}} -\int_{s_{max}/4}^{s_{max}}\right] 
ds_1 ds_2  f_1(s_1) f_2(s_2)
 \\
&=& \left[ F_1(s_{max})- F_1(0)\right]
    \left[ F_2(s_{max})- F_2(0)\right]
 \\
&&- \left[ F_1(s_{max})- F_1(\frac{s_{max}}{4})\right]
    \left[ F_2(s_{max})-
    F_2(\frac{s_{max}}{4})\right],
\;\;\;
\end{eqnarray}
\begin{equation}
 F_j(x) = \int^x  f_j(s)ds, \;\; j=1,2.
\end{equation}
%
\end{itemize}

Finally we make a choice of a suitable form of $f_j$ functions.
We use a multi-branch form of it, each branch $\alpha$ being
associated with a basic mass singularity type:
\begin{eqnarray}
 f_j(s) &=& 
   \sum_\alpha^{N_{MAS}} a_{j\alpha} 
      \frac{ f_{j\alpha}(s)}{\int_0^{s_{max}} f_{j\alpha}(x) dx},\;\;
\sum_\alpha^{N_{MAS}} a_{j\alpha} =1,
\\
 f_{j\alpha}(s) &=& \left\{ \begin {array}{l}
                1,\\ \\
   {(s+m_R^2)^{-n}},\\ \\
   \left(s+m_R^2\right)^{-1}
         {\log^n\left((s_{max}+m^2_R)/(s+m^2_R)\right)},\\ \\
   {\left((s-M^2_k)^2 +M^2_k\Gamma^2_k\right)^{-1}},\\ \\
   {\left(s_{max}+m_R^2-s\right)^{-1}}.
                        \end{array}
                \right.
\label{FMASS}
\end{eqnarray}
The $N_{MAS}$ denotes the total number of ``mass branches''. 
The parameter $m_R^2$ takes a role of a regulator, preventing distributions from
blowing up to infinity at the phase-space limit. 
It should be stressed that $m_R^2$ does not play the role of a cut-off, 
and that the complete phase space is covered by the generation. 
The $M_k$ and $ \Gamma_k$ are parameters of the $k$-th resonance.
All the functions presented above are easily integrable into the elementary
functions.


\subsection{Angular Distributions}


In the next step we generate the angular variables: $\cos\theta_j$ and
$\phi_j$ (the index $j=1,2,3$ denotes the number of angle
within the same branch).

The role of the angular variables is to describe the class of $t$- and
 $u$-channel-type
singularities of the Feynman graphs. There is a number of different
transfers that can be built, based on two beams and 
four external final-state four-vectors. 
In principle each of them can develop a singularity  in some of the
 Feynman graphs for some configurations
 of final-state fermions.

The general angular function 
$g(\cos\underline \theta,\underline \phi,\underline s)$ of 
eq.\ \reff{FACTORISATION} is further
factorized with respect to the angles of each step of the
generation. Unlike the mass distributions case, the factorization is
not complete -- each subsequent angle uses previously generated angles
and masses (in the form of four-momenta). We write it symbolically
as:
\begin{equation}
g(\cos\underline \theta,\underline \phi,\underline s)= 
   g_1(\cos\theta_1,\phi_1;\underline s)
   g_2(\cos\theta_2,\phi_2;\theta_1,\phi_1,\underline s)
   g_3(\cos\theta_3,\phi_3;\theta_1,\phi_1,\theta_2,\phi_2,\underline s).
\end{equation}

Much as with the mass distributions, we use branching structure to
build individual $g_j$ functions (we always take 
flat distributions in $\phi_j$ variables and thus their generation
decouples):
\begin{eqnarray}
 g_j(\cos\theta_j,\phi_j;...,\underline s) &=& 
    \sum_{\alpha}^{N_{ANG}} 
         b_{j\alpha}
         \frac{ g_{j\alpha}(\cos\theta_j,\underline s) }{
           \int_{-1}^{1} d\cos\theta \int_{0}^{2\pi} d\phi
            g_{j\alpha}(\cos\theta,\underline s)},\;\;
\sum_{\alpha}^{N_{ANG}} b_{j\alpha} =1,
\;\;
\end{eqnarray}
with $N_{ANG}$ denoting the number of ``angular branches''.
The basic angular distribution functions $g_{j\alpha}$ are:
\begin{eqnarray}
 g_{j\alpha}(\cos\theta_j,\underline s) &=& \left\{ \begin {array}{l}
    1,
\\ \\
   {\left(-t_j+m_A^2\right)^{-n}},
\\ \\
   \left(-t_j+m_A^2\right)^{-1}
       {\log^n\left((-t_{j,max}+m_A^2)/(-t_j+m_A^2)\right)},
\\ \\
   {\left(-u_j+m_A^2\right)^{-n}},
\\ \\
   \left(-u_j+m_A^2\right)^{-1}
       {\log^n\left((-u_{j,max}+m_A^2)/(-u_j+m_A^2)\right)},
                        \end{array}
                \right.
\label{GANGLE}
\end{eqnarray}
where $t_j$ and $u_j$ are functions of $\cos\theta_j$ and 
$\underline s$-variables. The
$\phi_j$ variables are independent and generated 
with a flat distribution. 
The $m_A^2$ is, as before, a small mass introduced to regularize the
distributions at singularity points.


\subsection{Event Construction}

Now, we can continue step by step with the generation and construction 
of the explicit form of the event with the help 
of the variables generated as explained in the previous sections.

Let us first define the relation between angles and $t_j$ and $u_j$
transfers.
For any two-to-two sub-process of our generation
(see Fig. 1), the transfers $t_j$ and $u_j$ are
defined as
\begin{equation}
t_j=(P_{1}-q_{1_j})^2,\;\;\;
u_j=(P_{1}-q_{2_j})^2,
\label{TUTRANSFERS}
\end{equation}
where $P_{1}$, $P_{2}$ denote four-momena  of the first
and second beams, and $q_{1_j}$, $q_{2_j}$ denote the four-momenta of the
first and second outgoing objects. These outgoing objects
can either be the final-state particles or groups of
particles (``intermediate states''). Later on, we will use the masses of the
objects to label them. Thus $[m_1^2]$, $[s_1]$ will denote respectively the
first final-state fermion and the system consisting of second, third and fourth 
particles (case of fig. 1a) or the 
system consisting of first and second particles 
(case of fig. 1b). In every  case, the invariant mass 
of the object is known, either from the input parameters or from the mass
generation in the previous step of the algorithm. In this
way one has an access to all possible transfers $t_j$ and
$u_j$.

Angles corresponding to the invariants are easiest to generate in the
local CMS frames of outgoing four-vectors $(q_{1_j},q_{2_j})$. 
In such a frame the $t_j(\cos\theta_j,\underline s)$ and 
$u_j(\cos\theta_j,\underline s)$ can
be expressed in a standard way
(we drop the index $j$ here):
\begin{eqnarray}
t &=& (P_1-q_1)^2 = q_1^2+P_1^2 -2q_1^0P_1^0 
                     +2\vert\vec q_1\vert\vert\vec P_1\vert\cos\theta,
\\
u &=& (P_1-q_2)^2 = q_1^2+P_2^2 -2q_1^0P_2^0 
                     -2\vert\vec q_1\vert\vert\vec P_1\vert\cos\theta,
\end{eqnarray}
with 
\begin{eqnarray}
q_1^0 &=& \frac{1}{2\sqrt{s_{12}}} \left(s_{12}+q_1^2-q_2^2\right),
\\
P_1^0 &=& \frac{1}{2\sqrt{s_{12}}} \left(s_{12}+P_1^2-P_2^2\right),
\\
\vert\vec q_1\vert^2 &=& \frac{1}{4}s_{12}
               \lambda\left(1,q_1^2/s_{12},q_2^2/s_{12}\right),
\\
\vert\vec P_1\vert^2 &=& \frac{1}{4}s_{12}
               \lambda\left(1,P_1^2/s_{12},P_2^2/s_{12}\right),
\\
s_{12} &=& (P_1+P_2)^2.
\end{eqnarray}

Finally we need to define a series of \ frames (i.e. four-momenta
of group of particles) used
to construct subsequent angles. This can be, as a matter of fact,
decoded from figs.~1a and 1b. 

For the configuration of fig.\ 1a we start by generating the angle
$\angle ([m_1^2],P_{1})$  
in the rest frame of the two beams 
$P_{1}$ and $P_{2}$.
Next, we subtract the generated final four-momentum
$[m_1^2]$ from the second beam $P_{2}$ and repeat the previous step for the
angle $\angle ([m_2^2],P_{1})$ in the rest frame of the pair $([m_2^2],[s_2])$. 
Finally, in the third step, after subtracting again
$[m_2^2]$ from $P_{2}-[m_1^2]$, we build the angle 
$\angle \left({[m_3^2]},{P_{1}}\right)$ in the rest frame of the pair
$\left({[m_3^2]},{[m_4^2]}\right)$.

In the case of fig.\ 1b, the first and second angles,
$\angle \left({[s_1]},{P_{1}}\right)$ in the rest frame of the pair 
$ \left({[s_1]},{[s_2]}\right)$ and
$\angle \left({[m_1^2]},{P_{1}}\right)$ in the rest frame of the pair
$ \left({[m_1^2]},{[m_2^2]}\right)$,
are generated as
before. In the definition of the third angle,
$\angle \left({[m_3^2]},{P_{2}}\right)$ in the rest frame of the pair
$\angle \left({[m_3^2]},{[m_4^2]}\right)$,
the direction of the 
second beam $P_{2}$ is chosen as the $z$-axis 
instead of the first one.


\subsection{Normalization}


The last step is to check out the normalization 
${\cal N}^i_{CR}$ of a given
$\psi^i_{CR}$ used in the definition of the branch $i$ of eq.\ \reff{TREE-NORM}. 
We need to calculate the integral
$\int d\Phi^i \psi_{CR}^i$:
\begin{eqnarray}
\label{NORM}
     {\cal N}^i_{CR} &=&
  \int\limits_{\Omega^i_{CR}} 
        d\Phi^i(\cos\underline \theta^i,\underline \phi^i,\underline s) 
     \psi_{CR}^i (\cos\underline \theta^i,\underline \phi^i,\underline s) 
\\
       &&    =
                    \int ds_1 ds_2 f(s_1,s_2)
   \int \prod_{j=1}^3 d\cos\theta_j d\phi_j 
   g_j(\cos\theta_1,\phi_1;\dots,\theta_j,\phi_j,\underline s)
\nonumber
       = 1.            
\end{eqnarray}
The normalization to unity follows from the specific form of the $f$ and $g$
functions of our choice.
   The property that normalization factors ${\cal N}^i_{CR}=1$, although
   not necessary for the correctness of the algorithm, simplifies the
   practical use. 
   Let us elaborate more on this point. Thanks to the normalization,
   and if in formulae \reff{FMASS} and \reff{GANGLE} only single branches 
   are present, the product of functions $f$ 
   and $g$ forms the Jacobian of the transformation from the 
   $\Omega^{i}_{CR}$ domain to the eight-dimensional unit cube. 
   The inverse transformation induces the measure on $\Omega^{i}_{CR}$, which is 
   the (crude) probability distribution. In the multichannel case,
   when coefficients $a_{j_\alpha}$ and  $b_{j_\alpha}$ can 
   all be non-zero, the measure (probability distribution) on $\Omega^{i}_{CR}$ 
   is defined as the weighted%
\footnote{
   More precisely, weighted with appropriate products 
   of coefficients/probabilities $a_{j_\alpha}$ and $b_{j_\alpha}$.
   } 
   sum of measures defined by individual transformations.


\subsection{Optimization}


Finally, we have to discuss the choice of the coefficients  
$p_i$, $a_{j\alpha}$ and $b_{j\alpha}$. It is these parameters that
assure flexibility of the algorithm.
The coefficients $p_i$, $a_{j\alpha}$ and $b_{j\alpha}$ are free
parameters of the generator. Their choice depends on the final state
chosen as well as on the external cuts used in the calculation of
cross-section. 
An optimal choice of these parameters is a delicate and important part of
the algorithm. We did it ``manually'', by looking at the
physical structures they approximate. Our strategy was to eliminate
the most overweighted events by iterative adjustments of these
coefficients and, if needed, adding more branches to the master sum of eq.\ 
\reff{TREE-BRANCH}. One can easily imagine other, more sophisticated,
optimization procedures, for example by assuming functional
dependence of these coefficients on the type of the final fermions or by
allowing for functional dependence on external cut-offs. This latter
optimization may be especially interesting. In the case of
strongly peaked cross-sections, strong cut-offs can downgrade the efficiency
of the algorithms by causing excessive rejection, unless compensated by 
retuning of these internal parameters.
Yet another option would be a numerical
optimization of some sort. One can mention in this context 
the approach of Ref.\ \cite{kleiss:1994}, for example.
 
To summarize, such a large number of free parameters makes it more
complicated to find the optimal configuration of the coefficients, but
on the other hand it gives a direct access to virtually {\em
every} singular configuration and gives us a possibility for its direct
modelling. 

We also have to remember that 
$p_i$, $a_{j\alpha}$ and $b_{j\alpha}$ are dummy
parameters. It means that, by varying them, one gets a
strong tool for testing the correctness of the integration.


\section{The Algorithm}


At this point we are ready to put together all the pieces and write down
the complete formula for the overall weight and normalization for our 
algorithm. To be able to calculate $\sigma$ of eq.\ \reff{SIGMA}, 
the following steps have to be completed:
\begin{enumerate}

\item
Generate the branch number $i$ according to probabilities $p_k$.

\item
Generate point 
$(\cos\underline \theta^i,\underline \phi^i,\underline s)$
according to the $\psi^i_{CR}$
distribution of the chosen $i$-th branch. 

\item
Construct the weight $w^a(i)$.

\item
If $w^a(i)\neq 0$ construct the phase-space point $(\underline q)$ out of
$(\cos\underline \theta^i,\underline \phi^i,\underline s)$.

\item
Calculate the total weight $w(i,P,\underline q)$ 
of eq.\ \reff{SIGMA3} (or \reff{SIGMA3-BIS}) in terms of
four-vectors.

\item
Calculate the cross-section as an average of the weight
$w(i,P,\underline q)$ 
\begin{equation}
\sigma = \left<w(i,P,\underline q)\right>.
\end{equation}

\end{enumerate}

A few comments are in order here.

While constructing a set of all branches,
one has to include all the necessary permutations
of external momenta of beams and outgoing fermions. 
The generic two branches 
--- ``chain'' and ``fork'', need to be applied to all configurations of 
external momenta to describe all the singularities properly. All together, 
this leads to over fifty branches in the case of four-fermion final states%
\footnote{
If one wanted to be even more precise, and counted separately 
each ``basic'' branch of eqs.\ \reff{FMASS} and \reff{GANGLE}, the total
number of branches would easily exceed $10^6$!
}.

The algorithm as described in this note is the one with variable weights.
By means of the standard rejection technique 
it can provide unweighted events as well.


\section{Summary}


In this paper we gave a complete description of the Monte Carlo
algorithm for the generation of the four-fermion phase space, as needed
for LEP2 applications. It is based
on the ``multichannel'' approach, as explained elsewhere e.g. in 
\cite{jadach:MC-www}.
Strong emphasis is given to the construction of separate 
branches with the help of rather simple modular building blocks.
This approach allowed for transparent writing of the code and testing,
as well as for easy extensions and modifications in the future. 

The major disadvantage of the present version, or rather the important 
place for future improvements,
is optimization of internal coefficients. In particular, in the 
case of big changes of external cut-offs or centre-of-mass energy, 
the coefficients may need retuning. At present this must be done
by hand.

The algorithm has been successfully implemented in the {\tt KORALW}
\cite{koralw:1998} Monte Carlo program. Generation can cover the full
phase space, or any sub-region defined by cut-offs\footnote{
In {\tt KORALW} some of the cut-offs can be optionally implemented 
before lengthy calculation of the matrix element. 
This is essential, e.g. in the generation
of $e^+e^- f \bar f$ final states with tagged electrons when the rejection 
rate is very high.
}. This is the case
of all four-fermion final states at LEP2 centre-of-mass energies.
Numerically stable results, with a statistical precision of few per mille,
can be obtained from the runs easily attainable in a few hours of CPU time
of any modern PC. 

\vskip 0.3 cm
\centerline{ \bf \large Acknowledgements}
\vskip 0.3 cm

Authors are indebted to S. Jadach, W. P\l{}aczek, A. Vallasi, M. Witek for 
stimulating discussions and cooperation at different steps of 
the algorithm development. One of the authors (MS) acknowledges
support of the CERN ALEPH group.

%


\end{document}